\documentclass[aps,prl,twocolumn,showpacs]{revtex4}
\usepackage{epsfig}
\usepackage{times}
\usepackage{amsmath}
\begin{document}

\title{Spontaneous Symmetry Breaking in Interdependent Networked Game}

\author{Qing Jin and Zhen Wang\footnote{Corresponding author: zhenwang0@gmail.com}}
\affiliation
{School of Physics, Nankai University, Tianjin 300071, China}

\begin{abstract}
Spatial evolution game has traditionally assumed that players interact with neighbors on a single network, which is isolated and not influenced by other systems. We introduce the simple game model into the interdependent networks composed of two networks, and show that when the interdependent factor $\alpha$ is smaller than a particular value $\alpha_C$, homogeneous cooperation can be guaranteed. However, as interdependent factor exceeds $\alpha_C$, spontaneous symmetry breaking of fraction of cooperators presents itself between different networks. In addition, our results can be well predicted by the strategy-couple pair approximation method.
\end{abstract}

\maketitle

It is widely recognized that cooperation is essential for the social and natural evolution. To support this issue, spatial structure plays a significant role in leading the emergence of cooperation via applying the social dilemma game \cite{SDG}. However, most of these studies are implemented on a single network which ignores several particular intimacies between social members in the real life, such as married couples or parents and children. In order to exactly delineate the real situations in the life some more sophisticated paradigms need to be presented, which could distinguish between the intimate and common relationships among people.

Recently, the property and function of interdependent networks regarding catastrophic cascade of failures have been investigated by Buldyrev {\it et al.} \cite{Innet1}. Rather than considering the catastrophic events on a single network, they provide the model on two interdependent networks where nodes on network A depend on network B, and vice versa \cite{Innet2}. They find that the interdependent networks are more vulnerable to random failures if they have broader degree distributions, which is contrary to the situation for one single network. The model of interdependent networks, in our views, becomes a perfect candidate to describe various types of human interactions. To be concrete, a player (a node) playing social dilemma game with his neighbors represents common relationship among human being, and the relation between him and his companion on the other network characterizes the specific intimate relationship like husband and wife, or father and son. In such a case, the player's decision is not simply dependent on his own payoff during the game but also relevant to his companion's situation. In this sense, the model provides a more realistic description of cooperation phenomenon by considering the combined effect of two of Nowak's cooperation mechanisms: network reciprocity and kin selection \cite{Nowak1}.

\begin{figure}
\begin{center}
\includegraphics[height=14cm,width=75mm]{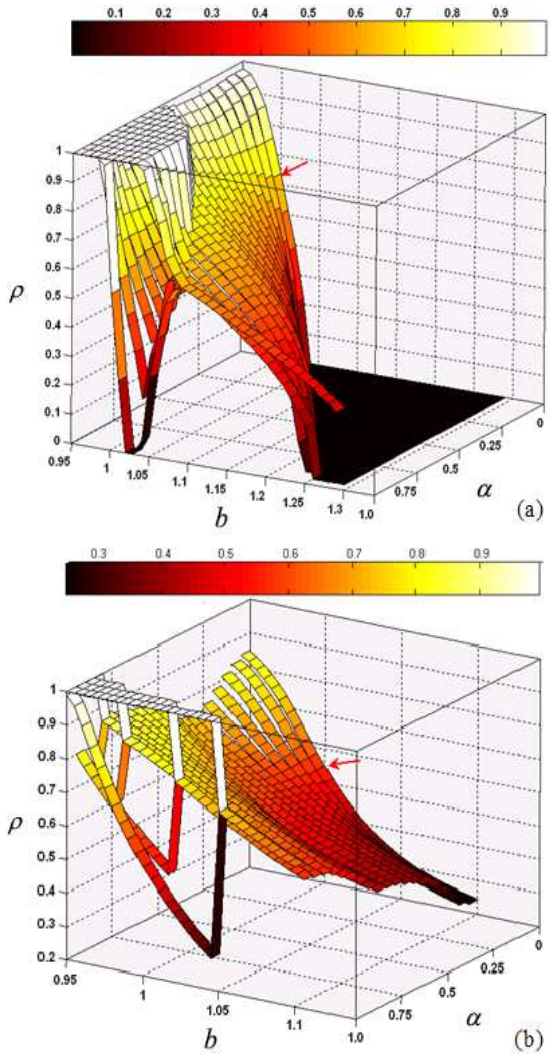}
\caption{(Color online) Fraction of cooperators $\rho$ as a function $b$ for different values of $\alpha$ when Monte Carlo simulations (a) are implemented. Note that if $b$ is larger than 1 (marked by the red arrows), faction of cooperators will enhance with the increment of $\alpha$. When $\alpha$ exceeds a specific value $\alpha_C$ ($\alpha_C \approx$0.5), the symmetry breaking phenomenon will appear. In addition, our strategy-couple pair approximation approach (b) correctly predicts the trends.}
\label{fig1}
\end{center}%
\end{figure}

In this Letter, we will perform the Prisoner's Dilemma Game (PDG), one of the most powerful models in studying cooperation phenomenon, on the interdependent networks (which are composed of Network Up and Network Down, see the {\it Appendix}). To control the intimacy between nodes from two networks, an interdependent factor $\alpha$ ($0 \le \alpha \le 1$) is proposed: in the limit $\alpha$$\to$0 two networks are refereed as weak interdependency; in the limit $\alpha$$\to$1 strong interdependency between networks will occur. We will show that, when $\alpha$ is smaller than a particular value $\alpha_C$, cooperation is highly promoted by setting a larger $\alpha$ and the distribution of cooperators is  homogeneous, which will be discussed below. However, when $\alpha$ exceeds $\alpha_C$, a spontaneous symmetry breaking between the fraction of cooperators on different networks can be observed. In order to analyze and explain these phenomena, we also extend the traditional pair approximation and give out the strategy-couple pair approximation (SCPA, see the {\it Appendix}).

As for the game, we will follow the Nowak-May framework, the so-called weak prisoner's dilemma game \cite{Nowak2}. Two players have a choice between two pure strategies, cooperators $C$ and defectors $D$. The payoffs are given by the following matrix:
\begin{eqnarray}
\begin{array}{cc}
\begin{array}{rr}\quad C& \quad D
\end{array}&\\
\begin{array}{l}C\\D
\end{array}
\left( \begin{array}{clr}
R & \quad S  \\
T & \quad P
\end{array}
\right)
\end{array}
\end{eqnarray}
$R$=1 is the reward for mutual cooperation, $T$=$b$ is temptation to defect, $S$=0 is the sucker's payoff and $P$=0 is the punishment for mutual defection, whereby $1\le b \le 2$ ensures a proper payoff ranking. In order to better characterize the influence about the interdependency of the networks, we choose the homogeneous network, including $L \times L$ regular lattices with periodic conditions and the small-world (SW) networks, since it is well-known that heterogeneous networks (such as scale-free network) would highly enhance cooperation in PDG \cite{hetnet}. Additionally, the interdependent networks also need to be point-to-point, which means every node in one network will have only one companion on the other network.

Each player located on the networks is initially designated to use either strategy $C$ or $D$ and acquires the payoff by playing the game with all his neighbors. Subsequently, one randomly chosen player $i$ selects one of his neighbors $j$ with equal probability, and adopts his strategy based on the Fermi Rule \cite{femir}: \begin{eqnarray}
W(i \to j)=f(G_j-G_i)=[1+exp(-(G_j-G_i)/K)]^{-1}
\end{eqnarray}
where $K$ represents the amplitude of noise (we simply fix $K$ to be 0.1 in this work ), and $G_i$ denotes the fitness of player $i$, considering both its own payoff $P_i$ and the payoff of its companion $P^{'}_i$. Of particular interest, the fitness $G_i$ can be quantitatively evaluated in the following way $G_i$=$(1-\alpha)*P_i$+$\alpha*P^{'}_i$. Here $0\le \alpha \le 1$ represents the interdependent factor. If $\alpha$=0 then the player's fitness is equal to its own payoff, and the model gets back to the original spatial PDG
\cite{orig}. If $\alpha$=1 then the fitness of a node will be totally determined by his companion's situation on the other network.

Results of Monte Carlo simulations presented below are obtained typically for the size $L$=100 to 400, but when we deal with the phase transition points, a large size ($L$=1000) will be adopted to assure the exactness of simulations. In a full Monte Carlo step (MCS) each players has a chance to adopt the strategy from one of their neighbors once on average. Moreover, the key quantity fraction of cooperators $\rho$ is determined with the last $10^4$ full steps of overall $5\times10^5$ MCS, and the final data results from an average over 20 independent realizations.

Figure 1 shows the simulation and analysis results about how fraction of cooperators $\rho$ varies as a function of $b$ for different values of interdependent factor $\alpha$. To give a clear illustration, the value of $\rho$ is also provided when $b$ is smaller than 1. As evidenced in the figure, we can observe two types of behaviors within the  system: symmetry breaking phenomenon and phase transition. When $\alpha$ is smaller than a particular value $\alpha_C$ ($\alpha_C$$\approx$0.5 in the present model), the fraction of cooperators will be completely equal on two networks. At the same time, it is worth emphasizing that with increasing interdependent factor $\alpha$ cooperation can  be better enhanced, which, to large extent, attributes to the self-organization of $C$-$C$ coupled clusters, as we will discuss in what follows. However, when $\alpha$ exceeds $\alpha_C$ the spontaneous symmetry breaking will emerge, namely, fraction of cooperators on two networks are different. In some particular regions (where $b$ is slightly larger than 1), all the players on one of the networks will uniformly choose the strategy $C$. With further raising the temptation to defection, the symmetry of the system will be gradually regained. In fact, the larger the value of $\alpha$, the more visible the symmetry breaking phenomenon. Moreover, Fig. 1(b) features the results of our SCPA approach (see the {\it Appendix}), which can correctly predict the trends. Because this analytical approach is unable to adequately provide the threshold value of cooperation (where the phase transition between mixed $C$+$D$ phase and pure $D$ phase occurs), we simply show the results with restricting $b$ between 0.95 and 1.10. From Fig. 1(b), we can see that SCPA method can perfectly predict the enhancement of cooperation when the larger $\alpha$ is considered, and it also qualitatively show us the emergence of the spontaneous symmetry breaking phenomenon \cite{sym}.

\begin{figure*}
\begin{center}
\includegraphics[height=3.25cm,width=157mm]{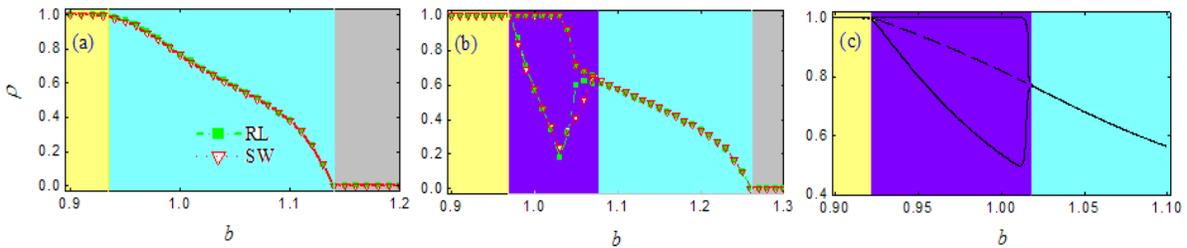}
\caption{(Color online) Phase diagrams for $\alpha$=0.4 (a) and $\alpha$=0.9 (b) on the interdependent regular lattices (green squares) and small-world (SW) networks (red triangles) with fraction of rewired links equalling 0.05. The colored regions represent different phases: yellow is pure cooperators phase (PC), cyan is mixed strategies phase (MS), gray is pure defectors phase (PD) and purple is symmetry breaking phase (SB). Additionally, the lines in (c) denote the results of strategy-couple pair approximation (SCPA) for $\alpha$=0.9, which is qualitatively similar to the case of (b). The dashed line is an unstable solution of SCPA approach.}
\label{fig1}
\end{center}%
\end{figure*}

In order to scrutinize the phase transition of system, now we turn to the phase diagrams
under different cases. Fig. 2(a) is the phase diagram of $\alpha$=0.4, which displays the existence of three phases: pure cooperators phase (PC), mixed strategies phase (MS) and pure defectors phase (PD), as was previously reported in the original PGD \cite{orig}. However, in Fig. 2(b), where $\alpha$ is set as 0.9 that is larger than $\alpha_C$, a novel phase emerges: the symmetry breaking phase (SB). In this phase, usually with one of the networks showing a pure-cooperation behavior, two networks do not share the same fraction of cooperators. The discovery of this interesting phase not only answers the question of why pure-cooperation can be found in human society, but also provides a key to the problem about the coexistence of pure-cooperation and quasi-cooperation in different aspects of our daily life. These seem reasonable and are easily justifiable with realistic examples. For example, when dealing with the commercial and military problems, we do not always choose cooperation as the strategy \cite{SDG}.  Whereas a pure-cooperation behavior is also ubiquitous in other aspects of natural and social lives, such as, the educational or academic events, collective behavior of ants or bees \cite{Nowak1}, where each member never betrays their groups, and the strength of society lies in the individual loyalty. We argue that if no additional rule is introduced, this pure-cooperation phenomenon goes beyond what can be supported by the traditional spatial reciprocity \cite{reci}. Moreover, it will be instructive to check the universality of this interesting behavior on other networks. From the presented results in Fig. 2(a) and Fig. 2(b), we find that the interdependent regular lattices (RL) and small-world (SW) networks actually share the same phase diagrams, implying that this behavior is robust to different coupled networks.

\begin{figure}
\begin{center}
\includegraphics[height=4.75cm,width=76mm]{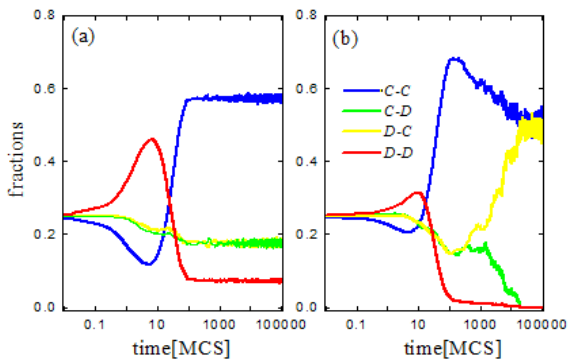}
\caption{(Color online) Time evolution of different strategy couples for $\alpha$=0.4 (a)
and $\alpha$=0.9 (b). About the meanings of couples, we can
give a simple example. For instance, $C$-$D$ means that the player on network Up chooses
$C$ and the interdependent player on network Down chooses $D$ (parameter: $b$=1.005).}
\label{fig1}
\end{center}%
\end{figure}

Importantly, the intriguing symmetry breaking phase can also be obtained by applying the SCPA approach (see Fig. 2(c)). Instead of the second phase transition in the  simulation results (Fig. 2(b)), SCPA shows a first order phase transition, namely, it cannot provide the exact type of phase transition. The reason of this shortage is that SCPA only considers two couples' interaction within the system, and neglects the long-range interaction among players on the networks, which actually plays an important role in the phase transition.  However, the flaw would not affect the prediction of SCPA about the system's behavior, such as,  showing us the symmetry breaking phase.  We also need to mention that there exists an unstable solution of the SCPA equations in the symmetry breaking phase (denoted by the dashed line), the details and equations of SCPA will be stated in the Appendix.

\begin{figure}
\begin{center}
\includegraphics[height=3.6cm,width=85mm]{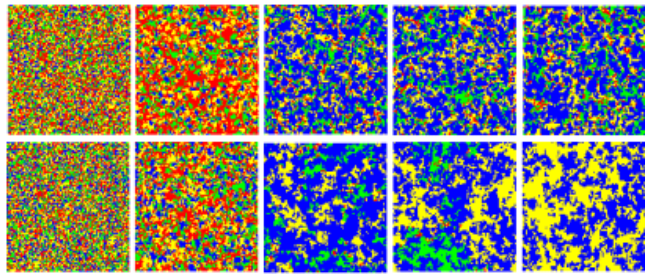}
\caption{(Color online) Evolution patterns of different couples for $\alpha$=0.4 (upper panel) and $\alpha$=0.9 (bottom panel). The color code of strategy couples is the same as Fig. 3, namely, $C$-$C$ blue, $C$-$D$ green, $D$-$C$ yellow and $D$-$D$ red. From left to right the specific steps are 0, 10, 200, 2000, and 30000 for both panels (parameter:
$b$=1.005).}
\label{fig1}
\end{center}%
\end{figure}

Subsequently, we proceed with examining the time evolution for four types of strategy couples: $C$-$C$, $C$-$D$, $D$-$C$ and $D$-$D$ couples. Figure 3 features the results obtained for $\alpha$=0.4 (a) and $\alpha$=0.9 (b), and the relevant evolution patterns are illustrated in Fig. 4. It is obvious that, in the very early stages of evolution process (note that fractions are recorded in between full steps), $D$-$D$ couples thrive. Quite surprisingly though, the tide changes fast, namely, $D$-$D$ couples become the rarest ones, and their dominant space is replaced by $C$-$C$ couples. However, in the next thousands of steps, the situations will become different within two systems (see Fig. 3). For $\alpha$=0.4 the system will reach the thermodynamic equilibrium state very quickly. When $\alpha$=0.9 is taken into account, an intriguing phenomenon appears: instead of
reaching the equilibrium state, clusters of $C$-$C$, $C$-$D$ and $D$-$C$ couples are gradually self-organized. In order to visually inspect this behavior, let us focus on the evolution patterns (see the bottom panel of Fig. 4). Initially, several sporadic clusters of $C$-$C$, $C$-$D$ and $D$-$C$ couples exist in the system, but soon they will combine to form larger clusters. Informed from the SCPA results (see Fig. 2(c)), we see that  the present pattern is probably unstable, which means that till now the system merely reaches a semi-equilibrium state and it cannot survive from little perturbations. Interestingly, this prediction comes true in the next steps: one type of the clusters dies out at last and only two kinds of them survive, which results in the spontaneous symmetry breaking of the networks since the fractions of $C$-$D$ and $D$-$C$ are not equal any more. The $C$-$C$
couples will remain in the system while the survival of $C$-$D$ or $D$-$C$
couples depends on the perturbations.

\begin{figure}
\begin{center}
\includegraphics[height=7.8cm,width=82mm]{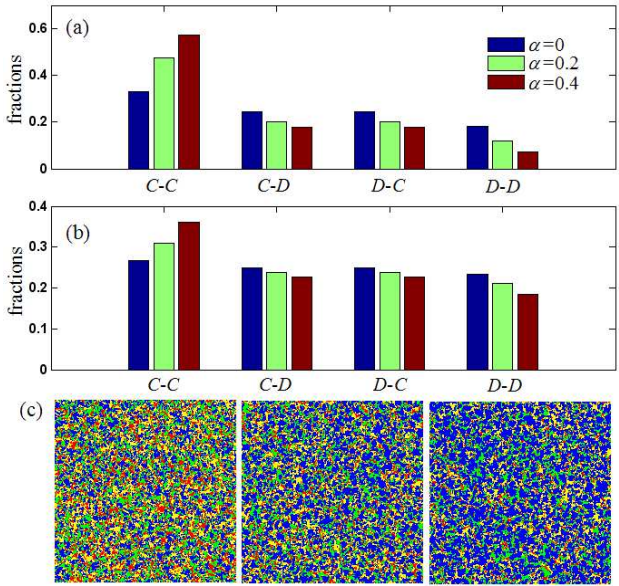}
\caption{(Color online) Fraction distributions of strategy couples for different value of $\alpha$ when simulation (a) and SCPA approach (b) are implemented. Simultaneously, we also show the spatial patterns in (c), where the color code is the same as Fig. 4
(parameter: $b$=1.005).}
\label{fig1}
\end{center}%
\end{figure}

Finally, it remains of interest to elucidate why cooperation can be improved with the increment of $\alpha$. To provide answers, we study the fraction distributions of strategy couples in Fig. 5. What firstly attracts our attention is the fact the larger the value of $\alpha$ is, the more $C$-$C$ couples exist. Actually, as increasing $\alpha$, there will be more $C$-$D$, $D$-$C$ and $D$-$D$ couples switching to $C$-$C$ couples (in the process, $D$-$D$ couples will first transform to $C$-$D$ or $D$-$C$ couples, then to $C$-$C$ couples). In addition, Fig. 4(c) shows the spatial patterns for different $\alpha$, whereby for $\alpha$=0 only a few sporadic $C$-$C$ couple clusters exist which comes from the occasional superposition of cooperators' clusters on both networks, since the networks are actually non-relevant in such a situation. However, when a larger $\alpha$ is considered ($\alpha$=0.2), more $C$-$C$ couples will be connected to each other in order to build solid clusters protecting themselves against the exploitation by defectors \cite{hetnet, femir, orig}. When $\alpha$ equals to 0.4 (close to the symmetry breaking value $\alpha_C$), $C$-$C$ couples strongly bond to each other, thereby much larger $C$-$C$ coupled clusters will be constructed in the system, which shows us a $C$-$C$ couples' ocean. At the same time, the $C$-$D$ and and $D$-$C$ couples sporadically exist through forming small clusters and $D$-$D$ couples can only survive along the edges of these small mixture strategy coupled clusters.

To conclude, we have introduced the interdependent networks into spatial game study. Through systematic simulations, we have demonstrated that the interdependency between different networks have a great influence on the cooperative
behavior. When the interdependent factor $\alpha$ exceeds a particular value $\alpha_C$,
the spontaneous symmetry breaking between the fraction of cooperators will appear. If it is smaller than $\alpha_C$, homogeneous fraction is able to be observed
in the system and the fraction will increase with the increment of $\alpha$. Besides, these phenomena could be well predicted and analyzed by our strategy-couple pair approximation
method.

This work is supported by the National Natural Science Foundation of China (Grant No. 10672081).

\end{document}


\title{Appendix: Strategy-Couple Pair Approximation Method}

\author{Qing Jin and Zhen Wang\footnote{Corresponding author: zhenwang0@gmail.com}}
\affiliation
{School of Physics, Nankai University, Tianjin 300071, China}

\maketitle

Before formal characterization of analysis method, it is very necessary to illustrate the structure of interdependent networks. For simplicity and without loss of generality, we only consider the case that interdependent networks are composed of two networks, Network Up and Network Down, both with the equivalent nodes (see Fig. A1). On each network nodes directly interact with their neighbors, as was previously depicted on the traditional network \cite{network1,network2,network3,network4,network5,network6,network7}. More importantly, the functioning of a node on one network also depends on the situation of corresponding node on the other network, denoted by the red dashed lines.

Next, we pay attention to the extended analysis approach. Pair approximation method
has proved its efficiency and importance to qualitatively predict the trends in the traditional spatial game study \cite{pap1,pap2,pap3,pap4}. However, we have extended the single network to interdependent networks, the method also need to be updated. Since we focus on the evolution of coupled strategies, the name $¡°$Strategy-Couple Pair Approximation(SCPA)$¡±$ may be more appropriate on interdependent networks.

Instead of considering different individuals on single network, we mainly deal with the couples containing players on both networks (see Fig. A2). Thus four types of couples will be found in the system: $C$-$C$, $C$-$D$, $D$-$C$ and $D$-$D$ couples. Notice the possible existence of spontaneous symmetry breaking phenomenon, the $C$-$D$ and $D$-$C$ couples should not be regarded as one single type. To keep their meanings more straightforward, we give a simple example. For instance, when the node on Network Up chooses $C$, and its corresponding one selects $D$. This case can be denoted by $C$-$D$. However, if the node on Network Up is $D$ and the interdependent one is $C$, $D$-$C$ is the correct notation for this situation. In addition, to keep expression simple, $C$-$C$, $C$-$D$, $D$-$C$ and $D$-$D$ will be represented by 1,2,3 and 4 in the following equations, respectively.

Based on all the above assumptions, when couple $I$ ($\left[I_u \atop I_d\right]$,  representing players on Network Up and Network Down) meets couple $J$($\left[J_u \atop J_d\right])$, the strategy adoption process can be written as two following equations,

\begin{eqnarray*}
p(\left[\begin{array}{cc}I_u&J_u\\I_d&J_d \end{array}\right] \to
  \left[\begin{array}{cc}J_u&J_u\\I_d&J_d \end{array}\right])&=&\sum_{X,Y,Z}\sum_{U,V,W} f(G_{J_u}-G_{I_u})\frac{p(X,I)p(Y,I)p(Z,I)p(I,J)p(J,U)p(J,V)p(J,W)}{{p_I}^3 {p_J}^3}\\
  \quad&=&\sum_{X,Y,Z}\sum_{U,V,W}u(I,J),\qquad(Aa1)
\end{eqnarray*}

\begin{eqnarray*}
p(\left[\begin{array}{cc}I_u&J_u\\I_d&J_d \end{array}\right] \to
  \left[\begin{array}{cc}I_u&J_u\\J_d&J_d \end{array}\right])&=&\sum_{X,Y,Z}\sum_{U,V,W} f(G_{J_d}-G_{I_d})\frac{p(X,I)p(Y,I)p(Z,I)p(I,J)p(J,U)p(J,V)p(J,W)}{{p_I}^3 {p_J}^3}\\
  \quad&=&\sum_{X,Y,Z}\sum_{U,V,W}d(I,J),\qquad(Aa2)
\end{eqnarray*}
where $p(S,S')$ represents the possibility of finding the couples of $S$ and $S'$, and $f$ is the transition probability (see Eq. (2)). The sum runs over all the possible types of neighbor couples for focal couple, namely,  $X$, $Y$, $Z$ for $I$, and $U$, $V$ , $W$ for $J$ (see Fig. A2). To make notations simple in what follows, the sum will be represented by $u(I,J)$ and $d(I,J)$. (Note that the functions of $u$ and $d$ not only depend on $I$ and $J$, but also rely on all their neighbor couples, the exact expressions of $u(I,J)$ and $d(I,J)$ should be $u(I,J)\vert_{X,Y,Z,U,V,W}$ and $d(I,J)\vert_{X,Y,Z,U,V,W}$). With respect to the individual fitness $G$, it is not only determined by his own payoff $P$, but also by the payoff of his companion $P'$ on the other network (see the text for details). While in this analysis method, the fitness of players belonging to couples $I$ and $J$ will be rewritten as follows,

\begin{eqnarray*}
G_{I_u}&=&(1-\alpha)*P_{I_u}+\alpha*P_{I_d},\\
G_{I_d}&=&(1-\alpha)*P_{I_d}+\alpha*P_{I_u},\\
G_{J_u}&=&(1-\alpha)*P_{J_u}+\alpha*P_{J_d},\\
G_{J_d}&=&(1-\alpha)*P_{J_d}+\alpha*P_{J_u}.\qquad\qquad\qquad\qquad\qquad\qquad(Aa*)
\end{eqnarray*}

If couple $J$ succeeds in impacting couple $I$, then the probabilities of various couple pairs will also change, which can result in 16 ordinary differential equations. Consider the symmetry of the equations and sum of the probabilities equalling to 1 (two elementary conditions for pair approximation method \cite{pap1,pap2,pap3,pap4}), the equations can be reduced as a set of 9 equations:

\begin{figure}
\begin{center}
\includegraphics[height=5.5cm,width=82mm]{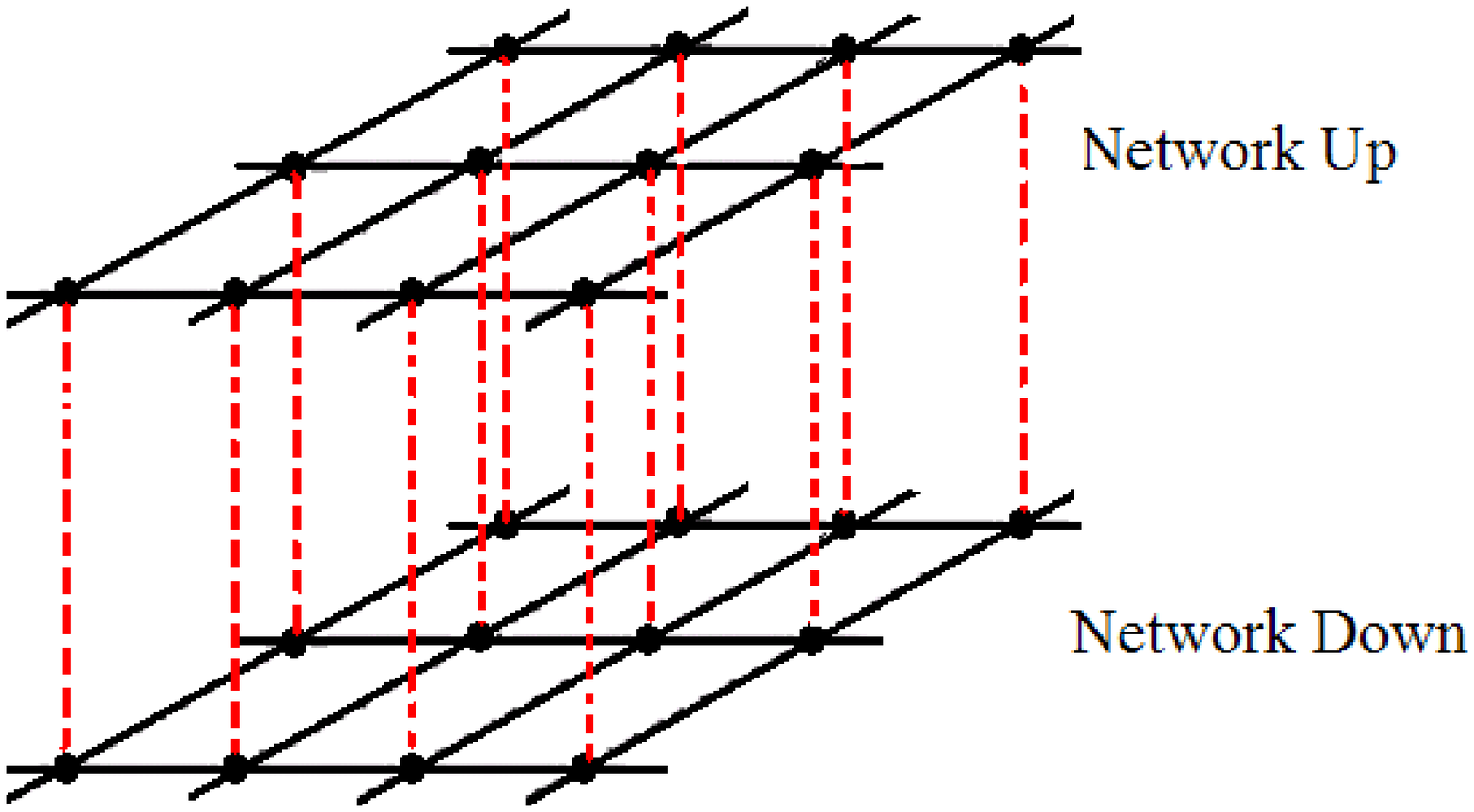}
\caption{(Color online) Illustration of two interdependent networks, Network Up and Network Down. On each network, the nodes not only interact with their neighbors, denoted by the black solid lines, but their situations also depend on the the corresponding nodes on the other network, represented by the red dashed lines. Here, we only consider the case of one-one correspondence between nodes on the networks.}
\label{fig. A1}
\end{center}%
\end{figure}

\begin{figure}
\begin{center}
\includegraphics[height=6cm,width=150mm]{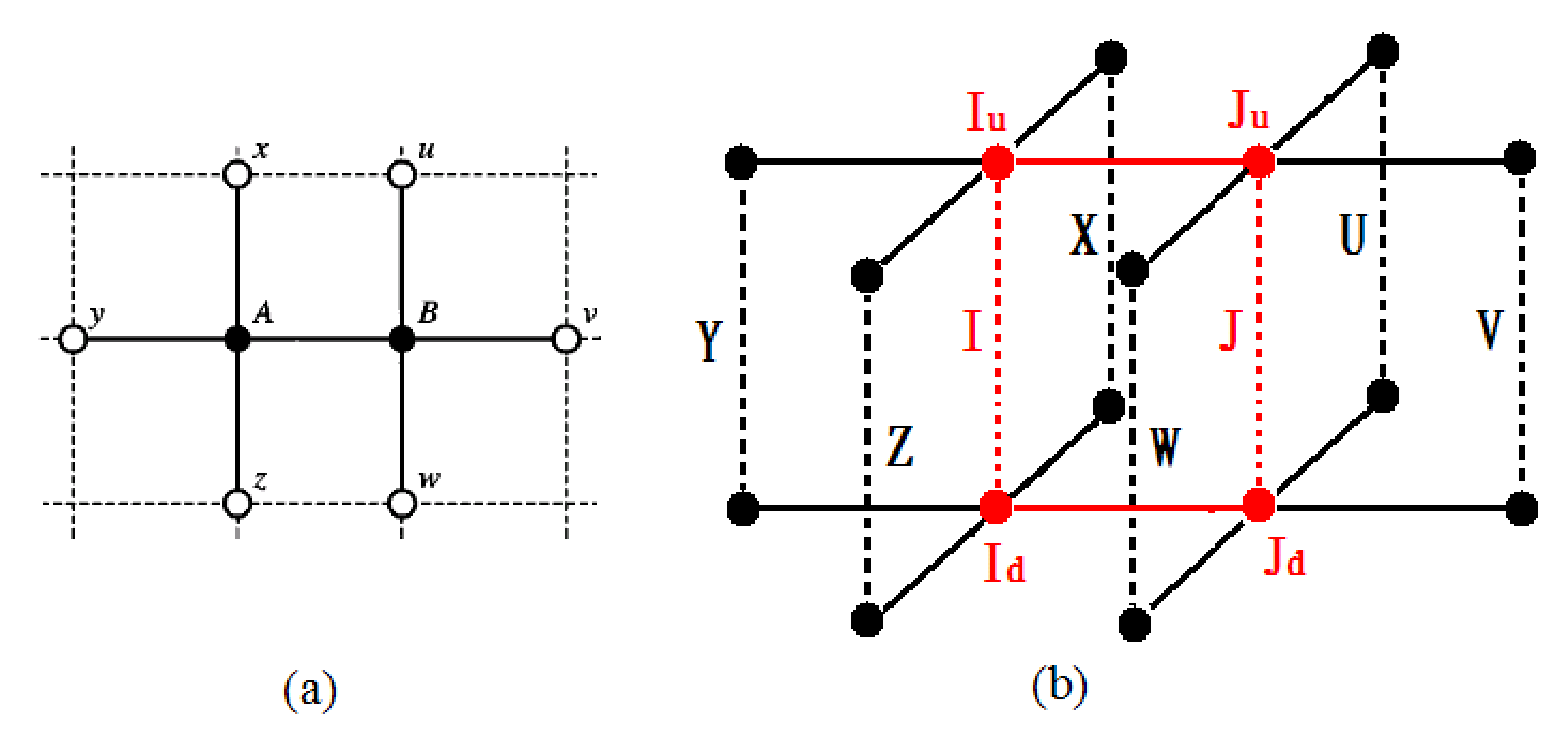}
\caption{(Color online) Demonstration of the relevant configurations for traditional pair approximation (a) and strategy-couple pair approximation (b). In (a), the configuration of nodes will determine the prediction results, since the probability of changing strategy fully depend on the situation of neighbor nodes on the single network \cite{pap1}. However, for the interdependent networks (b), the probability of changing strategy for focal couples (marked by red dashed lines) rely on their neighbor couples (denoted by black dashed lines). That is, $X$, $Y$, $Z$ are neighbor couples for couple $I$, and $U$, $V$, $W$ are neighbor couples for couple $J$.}
\label{fig. A1}
\end{center}%
\end{figure}

\begin{eqnarray*}
\dot{p}(1,1)=\sum_{X,Y,Z}\sum_{U,V,W}2[(e_1+1)p(1,3)u(3,1)-e_1p(1,3)u(1,3)-e_1p(1,4)u(1,4)+e_1p(2,3)u(3,2)]+{}
\nonumber\\
            {}\sum_{X,Y,Z}\sum_{U,V,W}2[(e_1+1)p(1,2)d(2,1)-e_1p(1,2)d(1,2)-e_1p(1,4)d(1,4)+e_1p(2,3)d(2,3)], \qquad(Ab1)
\end{eqnarray*}

\begin{eqnarray*}
\dot{p}(1,2)=\sum_{X,Y,Z}\sum_{U,V,W}(e_1+1)p(1,4)u(4,1)+(e_2+1)p(2,3)u(3,2)+e_2p(1,3)u(3,1)+e_1p(2,4)u(4,2)-{}
\nonumber\\
             {}-e_2p(1,3)u(1,3)-e_2p(1,4)u(1,4)-e_1p(2,3)u(2,3)-e_1p(2,4)u(2,4)+{}
\nonumber\\
            {}\sum_{X,Y,Z}\sum_{U,V,W}-(e_1+1)p(1,2)d(2,1)+e_2p(1,2)d(2,1)+e_1p(1,2)d(1,2)-(e_2+1)p(1,2)d(1,2)+{}
\nonumber\\
            {}+e_2p(2,3)d(2,3)+e_1p(1,4)d(1,4)-e_2p(1,4)d(1,4)-e_1p(2,3)d(2,3), \qquad(Ab2)
\end{eqnarray*}

\begin{eqnarray*}
\dot{p}(1,3)=\sum_{X,Y,Z}\sum_{U,V,W}-(e_3+1)p(1,3)u(1,3)-(e_1+1)p(1,3)u(3,1)+e_3p(1,3)u(3,1)+e_3p(2,3)u(3,2)+{}
\nonumber\\
             {}+e_1p(1,3)u(1,3)+e_1p(1,4)u(1,4)-e_3p(1,4)u(1,4)-e_1p(2,3)u(3,2)+{}
\nonumber\\
            {}\sum_{X,Y,Z}\sum_{U,V,W}(e_1+1)p(1,4)d(4,1)+(e_3+1)p(2,3)d(2,3)+e_3p(1,2)d(2,1)+e_1p(3,4)d(4,3)-{}
\nonumber\\
            {}-e_3p(1,2)d(1,2)-e_3p(1,4)d(1,4)-e_1p(2,3)d(3,2)-e_1p(3,4)d(3,4), \qquad(Ab3)
\end{eqnarray*}

\begin{eqnarray*}
\dot{p}(1,4)=\sum_{X,Y,Z}\sum_{U,V,W}-(e_4+1)p(1,4)u(1,4)-(e_1+1)p(1,4)u(4,1)+e_4p(1,3)u(3,1)+e_4p(2,3)u(3,2)+{}
\nonumber\\
             {}+e_1p(2,3)u(2,3)+e_1p(2,4)u(2,4)-e_4p(1,3)u(1,3)-e_1p(2,4)u(4,2)+{}
\nonumber\\
            {}\sum_{X,Y,Z}\sum_{U,V,W}-(e_4+1)p(1,4)d(1,4)-(e_1+1)p(1,4)d(4,1)+e_4p(2,3)d(2,3)+e_4p(1,2)d(2,1)+{}
\nonumber\\
            {}+e_1p(2,3)d(3,2)+e_1p(3,4)d(3,4)-e_4p(1,2)d(1,2)-e_1p(3,4)d(4,3), \qquad(Ab4)
\end{eqnarray*}

\begin{eqnarray*}
\dot{p}(2,2)=\sum_{X,Y,Z}\sum_{U,V,W}2[(e_2+1)p(2,4)u(4,2)-e_2p(2,3)u(2,3)-e_2p(2,4)u(2,4)+e_2p(1,4)u(4,1)]+{}
\nonumber\\
            {}\sum_{X,Y,Z}\sum_{U,V,W}2[(e_2+1)p(1,2)d(1,2)-e_2p(1,2)d(2,1)-e_2p(2,3)d(2,3)+e_2p(1,4)d(1,4)], \qquad(Ab5)
\end{eqnarray*}

\begin{eqnarray*}
\dot{p}(2,3)=\sum_{X,Y,Z}\sum_{U,V,W}-(e_3+1)p(2,3)u(2,3)-(e_2+1)p(2,3)u(3,2)+e_3p(2,4)u(4,2)+e_3p(1,4)u(4,1)+{}
\nonumber\\
             {}+e_2p(1,3)u(1,3)+e_2p(1,4)u(1,4)-e_3p(2,4)u(2,4)-e_2p(1,3)u(3,1)+{}
\nonumber\\
            {}\sum_{X,Y,Z}\sum_{U,V,W}-(e_3+1)p(2,3)d(2,3)-(e_2+1)p(2,3)d(3,2)+e_3p(1,2)d(1,2)+e_3p(1,4)d(1,4)+{}
\nonumber\\
            {}+e_2p(1,4)d(4,1)+e_2p(3,4)d(4,3)-e_3p(1,2)d(2,1)-e_2p(3,4)d(3,4), \qquad(Ab6)
\end{eqnarray*}

\begin{eqnarray*}
\dot{p}(2,4)=\sum_{X,Y,Z}\sum_{U,V,W}-(e_4+1)p(2,4)u(2,4)-(e_2+1)p(2,4)u(4,2)+e_4p(2,4)u(4,2)+e_4p(1,4)u(4,1)+{}
\nonumber\\
             {}+e_2p(2,3)u(2,3)+e_2p(2,4)u(2,4)-e_4p(2,3)u(2,3)-e_2p(1,4)u(4,1)+{}
\nonumber\\
            {}\sum_{X,Y,Z}\sum_{U,V,W}(e_4+1)p(1,4)d(1,4)+(e_2+1)p(2,3)d(3,2)+e_4p(1,2)d(1,2)+e_2p(3,4)d(3,4)-{}
\nonumber\\
            {}-e_4p(1,2)d(2,1)-e_4p(2,3)d(2,3)-e_2p(1,4)d(4,1)-e_2p(3,4)d(4,3), \qquad(Ab7)
\end{eqnarray*}

\begin{eqnarray*}
 \dot{p}(3,3)=\sum_{X,Y,Z}\sum_{U,V,W}2[(e_3+1)p(1,3)u(1,3)-e_3p(1,3)u(3,1)-e_3p(2,3)u(3,2)+e_3p(1,4)u(1,4)]+{}
\nonumber\\
            {}\sum_{X,Y,Z}\sum_{U,V,W}2[(e_3+1)p(3,4)d(4,3)-e_3p(2,3)d(3,2)-e_3p(3,4)d(3,4)+e_3p(1,4)d(4,4)], \qquad(Ab8)
\end{eqnarray*}

\begin{eqnarray*}
\dot{p}(3,4)=\sum_{X,Y,Z}\sum_{U,V,W}(e_4+1)p(1,4)u(1,4)+(e_3+1)p(2,3)u(2,3)+e_4p(1,3)u(1,3)+e_3p(2,4)u(2,4)-{}
\nonumber\\
             {}-e_4p(1,3)u(3,1)-e_4p(2,3)u(3,2)-e_3p(1,4)u(4,1)-e_3p(2,4)u(4,2)+{}
\nonumber\\
            {}\sum_{X,Y,Z}\sum_{U,V,W}-(e_4+1)p(3,4)d(3,4)-(e_3+1)p(3,4)d(4,3)+e_4p(1,4)d(4,1)+e_4p(3,4)d(4,3)+{}
\nonumber\\
            {}+e_3p(2,3)d(3,2)+e_3p(3,4)d(3,4)-e_4p(2,3)d(3,2)-e_3p(1,4)d(4,1), \qquad(Ab9)
\end{eqnarray*}
where the variable $e_l (l=1,2,3,4)$ represents the number of $l$ couples among $X$, $Y$, $Z$ (a more accurate demonstration should be $e_l$($X$,$Y$,$Z$)).